\documentclass[prl,amsmath,amssymb,twocolumn,showpacs,superscriptaddress]{revtex4-1}

\usepackage{bm}
\usepackage{amssymb}
\usepackage{colordvi}
\usepackage{graphicx}
\usepackage{color}
\usepackage{hyperref}
\usepackage{epsf}
\usepackage{subfigure}
\usepackage{epsfig,amsopn}

\newcommand{\be}{\begin{equation}}
\newcommand{\ee}{\end{equation}}
\newcommand{\bea}{\begin{eqnarray}}
\newcommand{\eea}{\end{eqnarray}}

\begin{document}

\title{Quantum efficiency bound for continuous heat engines coupled to non-canonical reservoirs} 
\author{Bijay Kumar Agarwalla}
\address{Department of Chemistry, University of Toronto, 80 Saint
 George Street, Toronto, ON, M5S 3H6, Canada}
\author{Jian-Hua Jiang}
\address{College of Physics, Optoelectronics and Energy, \&
Collaborative Innovation Center of Suzhou Nano Science and
Technology, Soochow University, 1 Shizi Street, Suzhou 215006,
China}
\author{Dvira Segal}
\address{Department of Chemistry, University of Toronto, 80 Saint
  George Street, Toronto, ON, M5S 3H6, Canada}

\date{\today}

\begin{abstract}
We derive an efficiency bound for continuous quantum heat engines absorbing heat
from squeezed thermal reservoirs. 
Our approach relies on a full-counting statistics description of nonequilibrium transport
and it is not limited to the framework of irreversible thermodynamics.
Our result, a generalized Carnot efficiency bound, is valid beyond the 
small squeezing and high temperature limit. 
%
%
Our findings are embodied in a prototype
three-terminal quantum photoelectric engine where a qubit converts heat absorbed
from a squeezed thermal reservoir into electrical power. We demonstrate that in
the quantum regime the efficiency can be greatly amplified by squeezing. 
From the fluctuation relation we further receive other operational measures in linear response, 
for example, the universal maximum power efficiency bound.
\end{abstract}

%
%


\maketitle

{\sl Introduction}.--- 
The efficiency of heat engines, defined by the ratio of the extracted
work to the absorbed heat, is fundamentally restricted by the second law of thermodynamics to the Carnot
limit.
%
This canonical bound is being challenged nowadays by quantum and classical effects \cite{review,kosloffrev}. 
For example, quantum phenomena such as steady state coherence  \cite{scully0, scully1, scully2}
and quantum correlations \cite{Huber}, which persist in multi-level quantum systems,
are suggested as a resource for the design of more efficient engines.
%
%
 %
%
As well, nonequilibrium, stationary reservoirs  
that are characterized by additional parameters besides 
their temperature, are exploited to construct devices with efficiency beyond the Carnot bound
\cite{Yi, LutzPRL, LutzEPL, Parrondo, David1, Kurizki15, Kurizki1, imamoglu}.  
%
%
In particular, a four-stroke Otto heat engine, operating between two reservoirs,
a hot {\it squeezed} thermal bath and a cold thermal bath,
was examined in Refs. \cite{Yi, LutzPRL, Parrondo}, reaching a unit value in the asymptotic, high squeezing limit.

Beyond the analysis of the averaged efficiency,
a quantum mechanical, full counting statistics derivation provides the ultimate, fundamental description of out-of-equilibrium
quantum statistical phenomena. Such an approach hands over symmetries, bounds, and noise terms (cumulants)
to characterize e.g. particle and energy transport.   
%
It is unclear, however, whether the steady state fluctuation symmetry \cite{fluct1,fluct2,fluct3} holds for transport phenomena
between non-canonical reservoirs. Another fundamental question is whether 
quantum principles impose new bounds on energy conversion
efficiency in such systems, to extend  the second law of thermodynamics. 

\begin{figure}[htbp] 
\vspace{-3mm} \hspace{-7mm}
{\hbox{\epsfxsize=65mm \epsffile{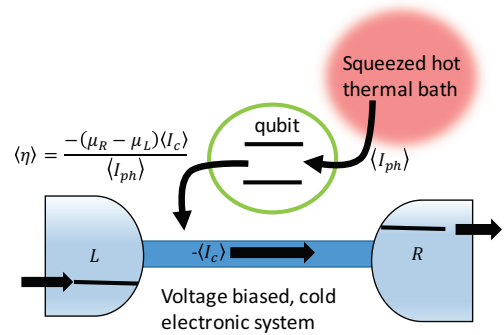}}}
\caption{Photoelectric quantum heat engine made of a qubit as the ``working fluid".  Energy absorbed  by the qubit from a hot 
squeezed thermal reservoir is converted to electrical power in the cold source-drain junction. }
\label{scheme}
\end{figure}

In this Letter, we fill these gaps by employing a full counting statistics approach to study energy conversion in quantum 
engines absorbing heat from a non-canonical reservoir.
Our device  consists a single qubit coupled to hot squeezed photon bath and two cold electronic reservoirs
 (the source and drain), see Fig. \ref{scheme}. We show that the nonequilibrium fluctuation relation (FR) for entropy
production can be recovered once identifying an effective temperature for
the squeezed thermal bath. From the fluctuation symmetry, we derive a
generalized, quantum efficiency bound for the heat engine, surpassing
the Carnot limit. Since the FR encompasses linear-response thermodynamics, we
receive immediately other operational measures of heat engines in
linear response: the universal maximum power efficiency bound
\cite{casati,espmp} and properties of fluctuations statistics \cite{espfl}. 
Our theory is exemplified with a quantum mechanical, full counting statistics 
description of a nanoscale photoelectric device.

%

We begin with a quick review of the fundamentals of the entropy production fluctuation theorem \cite{fluct1,fluct2,fluct3}.
Based on the microreversibility of the Hamiltonian dynamics and the canonical form of the initial condition,
one can prove a universal relation in steady state, 
%
\bea
{\rm ln} \left[ \frac{P_t(\Delta S)}{P_t(-\Delta S)}\right] = \Delta S.
\label{eq:PS}
\eea
Here, $P_t(\Delta S)$ is the probability distribution for entropy production $\Delta S$ during a time interval $t$. It is convenient to define the characteristic function
\bea
\mathcal{Z}(\lambda) \equiv \int d \Delta S \,e^{i\lambda \Delta S}\, P_t(\Delta S), 
\label{eq:Zdef}
\eea
with $\lambda$ the so-called counting parameter.
One can immediately prove the Gallavotti-Cohen fluctuation symmetry from the fluctuation relation (\ref{eq:PS}),
$\mathcal {Z}(\lambda)= \mathcal {Z}(-\lambda+i)$ \cite{fluct1,fluct2,fluct3}.
Moreover, by using $\lambda=0$ in Eq. (\ref{eq:Zdef}), it is easy to prove that $1 = \langle e^{-\Delta S}\rangle$.  
This equality immediately leads to the second law of thermodynamics, $\langle \Delta S\rangle \geq 
0$, by using Jensen's inequality for convex functions.

{\sl Three-terminal photoelectric devices}.---
We now apply these considerations onto a quantum heat engine consisting of
three terminals. In our construction, see Fig. \ref{scheme},
a qubit is coupled to a photonic heat source ($ph$), which may be canonical (equilibrium) or squeezed (out of equilibrium).
As well, the qubit is exchanging energy with an electronic circuit with two metal leads, $L$ and $R$,
which can be set out of equilibrium by the application of a finite voltage bias $\Delta \mu=\mu_R-\mu_L$  
and a temperature difference.
For simplicity, we assume that the two electrodes are maintained at the same temperature, 
$\beta_{el}=\beta_{\alpha}$; $\alpha=L,R$, and that the photon bath is hotter than the electronic system,
$\beta_{ph}<\beta_{el}$.
Our interest here is in the conversion of photon energy into electrical work.

In order to describe the system quantum mechanically, we use
 the two-time measurement protocol \cite{fluct1,fluct2} and define the characteristic function as
\bea
&&\mathcal{Z}(\lambda_c,\lambda_e,\lambda_{ph})
\nonumber\\
&&=\langle e^{i\lambda_c \hat A_c + i\lambda_e \hat A_e + i\lambda_{ph}\hat A_{ph}  }
e^{\!-i\lambda_c \hat A_c(t) \!- i\lambda_e \hat A_e(t) \!- i\lambda_{ph}\hat A_{ph}(t)  }
\rangle.
\nonumber
\label{eq:Z}
\eea
Here,  $\lambda_{c,e,ph}$ are counting parameters for charge, electronic energy, and photonic energy, respectively. 
$\hat A_c$, $\hat A_e$ and $\hat A_{ph}$ are the respective operators: 
$\hat A_c$ is the number operator corresponding to the total charge in e.g. the $R$ lead. 
$\hat A_e$ is the Hamiltonian operator for the $R$ electrode and
$\hat A_{ph}$ is the Hamiltonian operator for the photon bath.
Time evolution corresponds to the Heisenberg representation,
 $\langle ... \rangle$ represents an average with respect to the total initial density matrix, 
which takes a factorized form with respect to the system ($s$) and ($L$, $R$ and $ph$) baths, 
$\rho_T(0)=\rho_s(0)\otimes \rho_L\otimes \rho_R\otimes \rho_{ph}$.
The state of the metal leads is described by a grand canonical distribution,
 $\rho_{\alpha}\!=\! \exp[-\beta_{el} (\hat H_{\alpha}-\mu_{\alpha} \hat N_{\alpha})]/Z_{\alpha}$,
with $Z_{\alpha}\!=\!{\rm Tr}\Big[ \exp[-\beta_{el} (\hat H_{\alpha}-\mu_{\alpha} \hat N_{\alpha})]\Big]$ 
as the partition function. 

{\sl Equilibrium thermal photon bath}.---
Let us begin by assuming that the state of the photon bath is canonical, 
$\rho_{ph}\!=\! \exp[-\beta_{ph} \hat H_{ph}] / Z_{ph}$,
with $Z_{ph}={\rm Tr}\big[\exp(-\beta_{ph} \hat H_{ph})\big]$.
%
The fluctuation relation (\ref{eq:PS}) translates to

\bea
\frac{P_t(N,E_e,Q_{ph})}{P_t(-N,-E_e,-Q_{ph})} =  e^{\beta_{el} \Delta \mu N + (\beta_{el}-\beta_{ph})Q_{ph}}.
\eea
Here, $N$ denotes the number of electrons transferred from $R$ to $L$ during the time interval $t$. Similarly,
$E_e$ is the electronic energy and $Q_{ph}$ photonic heat that are 
exchanged between the baths during the time interval $t$.
The characteristics function thus satisfies
\bea
&&\mathcal Z(\lambda_c,\lambda_e,\lambda_{ph})
\nonumber \\
\!&&= \mathcal Z\big(\!-\lambda_c \!+ i\beta_{el}(\mu_R\!-\mu_L),-\lambda_e,\!-\lambda_{ph}\!-i(\beta_{ph}\!-\!\beta_{el})\big)
\label{eq:Zth}
\eea
This  relation immediately implies that
%
\bea
1= \langle e ^{-\beta_{el} \Delta \mu  N + (\beta_{ph}-\beta_{el})Q_{ph} }\rangle.
\eea
Using Jensen's inequality, we receive
$ [-\beta_{el} \Delta \mu \langle N \rangle + (\beta_{ph}-\beta_{el})\langle Q_{ph}\rangle ] \leq 0$.
The efficiency,
$\langle \eta\rangle \equiv -\frac{\Delta \mu \langle N\rangle }{ \langle Q_{ph}\rangle}$,  \cite{comment}
thus obeys the Carnot bound ($T=1/\beta$),
\bea
\langle \eta \rangle \leq \frac{\beta_{el}-\beta_{ph}}{\beta_{el}} =1 -\frac{T_{el}}{T_{ph}}.
\label{eq:eta}
\eea

{\sl Non-canonical photon bath}.---
We now repeat this exercise---with a squeezed, hot thermal reservoir.
The electric field of a single-mode wave
can be written as a combination of orthogonal (quadrature) components, 
which oscillate as $\cos \omega t$ and $\sin \omega t$ \cite{QObook}.
Squeezed states have reduced fluctuations in one of the quadratures---but enhanced noise
in the other quadrature---so as to satisfy the bosonic commutation relation. 
Such states are defined by two parameters, the squeezing factor $r$ and phase $\phi$ \cite{QObook}.

For simplicity, the quantum ``working fluid" system includes a single qubit with an energy gap $\hbar \omega_0$.
The squeezed bath can excite and de-excite the qubit, with rate constants $k_{u}^{ph}$ and $k_{d}^{ph}$,
satisfying \cite{Yi}
\bea
\frac{k_d^{ph}}{k_u^{ph}}= \frac{N(\omega_0)+1} {N(\omega_0) }. 
\label{eq:db}
\eea
Here \cite{breuer},
$N(\omega_0)= N_{th}(\omega_0) \left( \cosh^2 r + \sinh ^2 r\right) + \sinh^2 r$,
%
with the squeezing parameter $r$ reflecting the nonequilibrium nature of the bath.  
The phase $\phi$
does not appear in this expression, as it only affects transients.
In fact, at weak system-photon bath coupling, this effective temperature describes as well harmonic systems. 
 For a canonical thermal bath ($r=0$), the occupation number reduces
to the Bose-Einstein distribution function, $N(\omega_0)\rightarrow N_{th}(\omega_0)=1/[e^{\beta_{ph}\hbar \omega_0} -1]$,
and the rate constants satisfy the detailed balance relation with respect to the photon bath,  
$k_d^{ph}(\omega_0)/k_u^{ph}(\omega_0) = e^{\beta_{ph}\hbar\omega_0}$.
To restore the detailed balance relation for  the  $r\neq 0$ case,
one can identify an effective temperature, which is unique in the present model \cite{Yi},
\bea
\beta_{eff} (\beta_{ph},r,\omega_0) = \frac{1}{\hbar\omega_0} \ln \frac{1+N(\omega_0)}{N(\omega_0)}.
\eea
Simple manipulations provide 
\bea
\beta_{eff}=
\beta_{ph}
+ \frac{1}{\hbar\omega_0}\ln\left[ \frac{ 1 + (1+e^{-\beta_{ph}\hbar\omega_0})\sinh^2r   }{ 1 + (1+e^{\beta_{ph}\hbar\omega_0})\sinh^2r} \right].
\label{eq:betaeff}
\eea
%
It is important to note that: (i) $\beta_{eff} \leq \beta_{ph}$.
This observation implies that
more work can be extracted from a squeezed bath, than the case with $r=0$.
(ii) The effective temperature (\ref{eq:betaeff}) may depend on system parameters, the energy gap $\omega_0$ in the present case.
However, in the small $r$ and high temperature limit one recovers a proper, ``thermodynamical"
temperature
\bea
\beta_{eff} \rightarrow \frac{\beta_{ph}}{1+2\sinh^2 r},
\label{eq:beffc}
\eea
which is solely described in terms of bath parameters. Therefore, in this limit
universal relations of traditional linear irreversible thermodynamics hold. 

Identifying the entropy production associated with the photon energy flow by
$\langle \Delta S\rangle=(\beta_{el}-\beta_{eff})\langle Q_{ph}\rangle$, 
we perform a quantum mechanical, counting statistics analysis, similarly to the canonical case,
and confirm the symmetry Eq. (\ref{eq:Zth}), only replacing $\beta_{ph}$ by $\beta_{eff}$,
\bea
&&\mathcal Z(\lambda_c,\lambda_e,\lambda_{ph}) \nonumber \\
&&\!= \mathcal Z(\!-\lambda_c \!+ i\beta_{el}(\mu_R\!-\!\mu_L),
\!-\lambda_e,\!-\lambda_{ph}\!-i(\beta_{eff}\!-\beta_{el})).
\eea
The FR implies that
$1= \langle e ^{-\beta_{el} \Delta \mu N + (\beta_{eff}-\beta_{el})Q_{ph}}\rangle$,
thus the averaged efficiency, 
$\langle \eta\rangle \equiv -\Delta \mu\langle N\rangle/\langle Q_{ph}\rangle $, 
is bounded by
\bea
\langle \eta \rangle \leq 1 - \frac{\beta_{eff}}{\beta_{el}}.
\eea
This bound is universal, holding beyond the squeezed-bath case. It is valid for any nonequilibrium
 thermal bath that can be characterized by a unique, stationary, effective temperature, see Ref. \cite{David1}  for some examples.
Explicitly, the efficiency bound for our photoelectric engine is given by
\bea
\langle \eta \rangle \leq 1 \!-\! \frac{T_{el}}{T_{ph}} \!+\! 
 \frac{1}{\beta_{el}\hbar\omega_0}\!\ln\left[ \frac{ 1 \!+\! (1+e^{\beta_{ph}\hbar\omega_0})\sinh^2r   }{ 1 \!+\! (1+e^{-\beta_{ph}\hbar\omega_0})\sinh^2r} \right]\quad
\label{eq:boundS}
\eea
which is the main result of our work. 
It was derived from the fluctuation theorem, and it is valid to describe continuous quantum heat engines,
unlike earlier studies, which were focused on four-stroke engines,
see e.g. Ref. \cite{David1}.
Since the third term in this expression is positive for nonzero $r$, squeezing of a thermal bath always increases the
heat-to-work efficiency bound. 

We now discuss several interesting limits of Eq. (\ref{eq:boundS}).
First, we expand it close to thermal equilibrium assuming $\sinh^2 r$ is a small parameter. 
As well, we assume that the temperature of the photon bath is high, $\beta_{ph}\hbar\omega_0\ll1$.
The expression in the square brackets  reduces to 
\bea
&& \ln\left[ 1+ 
 \frac{ \left(e^{\beta_{ph}\hbar\omega_0}  - e^{-\beta_{ph}\hbar\omega_0}\right)  \sinh^2r   } { 1 + (1+e^{-\beta_{ph}\hbar\omega_0})\sinh^2r} 
\right] 
\nonumber\\
&&\rightarrow \frac{\beta_{ph}\hbar\omega_0 \times 2\sinh^2 r}{1+ 2\sinh^2 r},
\eea
and Eq. (\ref{eq:boundS}) becomes 
\bea
\langle \eta \rangle \leq 1- \frac{T_{el}}{T_{ph} (1+2\sinh^2r)}.
\label{eq:lutz}
\eea
Remarkably, this agrees with
Ref. \cite{LutzPRL,Parrondo}. Recall that our derivation concerns continuous heat engines;
Refs. \cite{LutzPRL,Parrondo}, in contrast,
received this limit by constructing a four-stroke cycle.
This agreement can be rationalized by noting that Eq. (\ref{eq:lutz}) should be regarded as 
a linear response limit for $r$, which 
is a resource to drive energy current between equal-temperature baths \cite{Yi}. 

\begin{figure}
\vspace{-1mm}
\includegraphics[width=80mm]{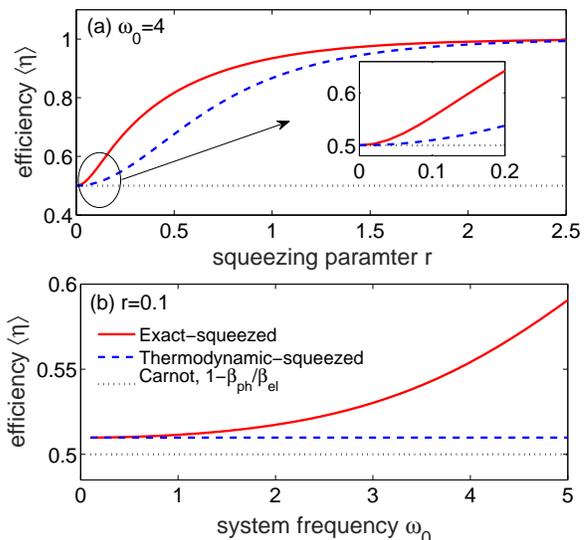} 
\vspace{-10mm}
\caption{
Efficiency bound as a function of (a) squeezing parameter
(b) subsystem frequency.
Exact result from  Eq. (\ref{eq:boundS}) (full), 
thermodynamical limit from Eq. (\ref{eq:lutz}) (dashed),
Carnot bound (dotted).
We use $\beta_{el}=2$ and $\beta_{ph}=1$.
}
\label{Fig2}
\end{figure}

Another interesting case is the
deep quantum regime, $\beta_{ph}\hbar\omega_0\gg 1$. 
Assuming small $r$, we receive from Eq. (\ref{eq:boundS})  an exponential quantum enhancement 
in comparison to the classical case,
\bea
\langle \eta\rangle \leq 1- \frac{T_{el}}{T_{ph}} + \frac{1}{\beta_{el}\hbar \omega_0} 
\left[ 
\frac{\sinh^2r}{1+\sinh^2r}  \times e^{\beta_{ph}\hbar \omega_0}
\right].
\eea
Note that the expansion assumes that the term inside the square bracket is kept below 1. 
Finally, at large $r$, the 
natural logarithm term in (\ref{eq:boundS}) cancels out the second contribution for both high and low $T_{ph}$. The
efficiency bound then saturates to a unit value, $\langle \eta\rangle \rightarrow 1$, 
realizing a complete conversion of heat to work.
We display these results in  Fig. \ref{Fig2}: Squeezing enhances the efficiency beyond the Carnot limit.
In the quantum regime, $\beta_{ph}\omega_0>1$, 
the bound is greatly reinforced beyond the ``thermodynamical" value, Eq. (\ref{eq:lutz}). 

A squeezed bath coupled to a qubit can be described by a 
single-unique effective temperature in the thermodynamical limit of high $T_{ph}$ and small $r$.
Since the fluctuation theorem embodies linear irreversible thermodynamics, 
all linear response operational results immediately follow.
In particular, the averaged maximum power efficiency (MPE) satisfies the universal linear response result \cite{casati} 
$\langle \eta^*\rangle =\langle \eta_M\rangle/2$, with $\langle \eta_M\rangle$ the upper bound in Eq. (\ref{eq:lutz}).
For a four-stroke Otto engine, the MPE is given by the Curzon-Ahlborn bound (beyond linear response),
$\langle \eta^*\rangle =1-\sqrt{\frac{T_{cold}}{T_{eff}}}$, 
with the identification of the thermodynamic temperature (\ref{eq:beffc}).
This agrees with Ref. \cite{LutzPRL}.

\begin{figure}
\hspace{4mm}
\includegraphics[width=75mm]{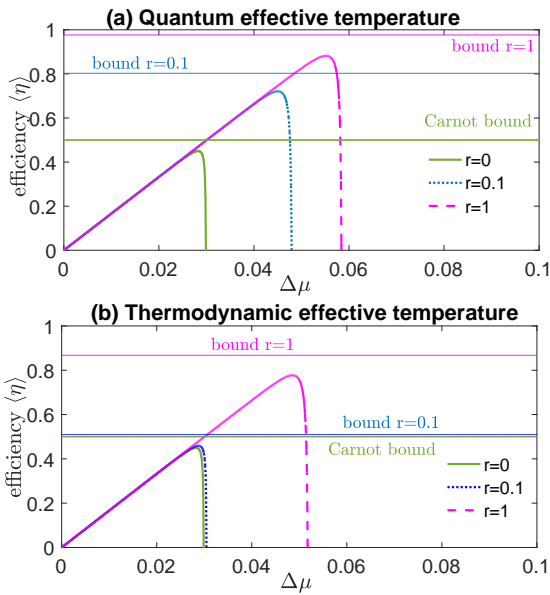}
\caption{
Efficiency of a photoelectric heat engine. Comparison of results making use of the
(a) quantum effective temperature (\ref{eq:betaeff}), (b) thermodynamical effective temperature 
(\ref{eq:beffc}).
Parameters are $\epsilon_d=-0.03$, $\epsilon_a=0.03$, $\omega_0=0.06$, 
$g=0.1$, $\Gamma_{L,R}=0.01$, $\Gamma_{ph}=0.1$, all in eV,
$T_{el}=30$ K, $T_{ph}=60$ K.   
The horizontal lines correspond to bounds, Eqs. (\ref{eq:boundS}) and (\ref{eq:lutz}). 
}
\label{Fig3}
\end{figure}

{\sl Example}.---
So far, we derived  an efficiency bound for continuous quantum heat engines based on the fluctuation symmetry. 
We now proceed and describe a device where a closed-form expression for the cumulant generating function (CGF)
${\cal G}(\lambda) = \lim_{t \to \infty} \frac{1}{t} \ln {\cal Z} (\lambda)$ is achieved.
Here, $\lambda$ collectively refers to the three counting fields. 
From the CGF, all cumulants of the charge current, electronic energy current 
and photonic current are available.
The closed-form expression for the efficiency of the engine allows us to 
examine its actual  performance under different conditions. 
%
Our model photoelectric heat engine is described by the Hamiltonian 
\bea
\hat H=\hat H_{s}+\hat H_{el}+\hat H_{ph} + \hat V_{s-el}+ \hat V_{s-ph}.
\eea
It comprises a single qubit $\hat H_s=\frac{\hbar \omega_0}{2}\hat \sigma_{z}$ of energy gap $\hbar \omega_0$.
The photon bath is written in terms of bosonic creation $\hat a_k^{\dagger}$ and annihilation $\hat a_k$ operators,
$\hat H_{ph}= \sum_{k}\omega_k \hat a_{k}^{\dagger}\hat a_k$. 
The electronic circuit includes two sites (quantum dots) denoted by 'd' and 'a', each coupled to their respective metal leads,
$L$ and $R$. The corresponding Hamiltonian is 
\bea
\hat H_{el}&=&  \epsilon_d \hat c_d^{\dagger}\hat c_d + \epsilon_a \hat c_a^{\dagger}\hat c_a + \sum_{\alpha,j}\epsilon_{\alpha,j}\hat c_{\alpha,j}^{\dagger}\hat c_{\alpha,j}  
\nonumber\\
&+&
\sum_{j}v_{L,j}\hat c_{L,j}^{\dagger}\hat c_d +  \sum_{j}v_{R,j}\hat c_{R,j}^{\dagger}\hat c_a + h.c.
\eea
Here $\hat c$  ($\hat c^{\dagger}$) are fermionic annihilation (creation) operators.
Energy is exchanged between the qubit and the reservoirs via the interaction terms
\bea
\hat V_{s\!-\!el}\!=\! g\hat \sigma_x\left (\hat c_d^{\dagger}\hat c_a \!+\!\hat  c_a^{\dagger}\hat c_d\right),
\hat V_{s\!-\!ph}\!=\! \hat \sigma_x\sum_{k}g_k\left(\hat a_{k}^{\dagger}\!+\!\hat a_k\right). \,\,
\eea
In words, the excitation or relaxation of the qubit couples to exchange of electrons between the two sites and the displacement
of harmonic modes.
The CGF is derived using a quantum master equation that is correct to second order in the
electron-qubit and the photon-qubit couplings \cite{PCCP, Bijay15},
\bea
{\cal G}(\lambda)
=
-\frac{1}{2}(k_{u} + k_{d}) + \frac{1}{2}\sqrt{(k_{u} - k_{d})^2 + 4\, k_{u}^{\lambda} k_{d}^{\lambda}}.
\label{eq:CGF-AH}
\eea
Here, $k_{d,u}^{\lambda}$ are the relaxation $(d)$ and excitation $(u)$ rate constants of the qubit, with transitions induced by the reservoirs,
e.g.,
\bea
k_d^{\lambda}\!=[k_d^{el}]^{\lambda}\!+\![k_d^{ph}]^{\lambda}, \,
[k_d^{el}]^{\lambda}\!=\! [k_d^{\lambda}]^{L\rightarrow R}\!+\! [k_d^{\lambda}]^{R\rightarrow L}\,\,.
\eea
%
Specifically, $[k_{d}^{\lambda}]^{L \to R}$ describes a de-excitation process of the qubit,
induced by an electron moving from the $L$ to the $R$ metal. It involves
 the release of energy at the right metal (where counting is performed), 
\bea
&&[k_{d}^{\lambda}]^{L \to R} =
\int \frac{d\epsilon}{2\pi} \Big[ f_L(\epsilon) (1-f_R(\epsilon+\omega_0)) J_L(\epsilon) J_R(\epsilon+\omega_0)
\nonumber\\
&&\times e^{-i(\lambda_c + (\epsilon + \omega_0)\lambda_e)} \Big].
\label{eq:rateint}
\eea
Here, e.g., $J_{L}(\epsilon)=g\frac{\Gamma_{L}(\epsilon)}{(\epsilon-\epsilon_d)^2 + \Gamma_L(\epsilon)^2/4}$ is the spectral
function of the $L$ metal, determined by the dot-metal hybridization energy
$\Gamma_{\alpha}(\epsilon)=2\pi\sum_{j}|v_{\alpha,j}|^2\delta(\epsilon-\epsilon_{\alpha,j})$.
Transitions induced by the squeezed photon bath satisfy \cite{PCCP} 
\bea
&&[k_{d}^{\lambda}]^{ph} =\Gamma_{ph} (\omega_0)\,[N(\omega_0)+1]\,e^{-i\lambda_{ph}\omega_0}, 
\nonumber\\
&&[k_{u}^{\lambda}]^{ph} =\Gamma_{ph}(\omega_0)\,N(\omega_0)\,e^{i\lambda_{ph}\omega_0}.
\eea
Here, $\Gamma_{ph}(\omega)=2\pi\sum_{k}|g_k|^2\delta(\omega-\omega_k)$, $N(\omega_0)$ was defined below Eq. (\ref{eq:db}).
It can be shown that the CGF (\ref{eq:CGF-AH}) obeys the FR.
%
%
The electron charge current is given by
\bea
\langle I_c\rangle = \frac{\partial \cal G(\lambda)}{\partial(i\lambda_c)} \Big|_{\lambda=0}
=\frac{ k_d\frac{\partial(k_u^{\lambda})}{\partial (i\lambda_c)} 
+ k_u\frac{\partial(k_d^{\lambda})}{\partial (i\lambda_c)}}{k_u+k_d}.
\eea
An analogous  expression is written for $\langle I_{ph}\rangle$. 
In Fig. \ref{Fig3} we display the averaged
 efficiency of the engine
$\langle \eta\rangle = -\Delta \mu \langle I_c\rangle / \langle I_{ph}\rangle$ for certain parameters,
once we set $T_{ph}>T_{el}$ and $\mu_R>\mu_L$.
The device operates as a photoelectric engine when heat is absorbed from the photon bath
and charge current is flowing against the potential bias. 
We operate it in the quantum regime, $\omega_0 \beta_{ph}\sim 10$, 
and reveal a significant enhancement of efficiency, largely exceeding the Carnot bound for small squeezing, $r=0.1$.

{\sl Summary}.---
We investigated the operation of heat engines coupled to a squeezed thermal bath. Based on the fluctuation symmetry, 
we derived a generalized quantum Carnot efficiency bound, as well as other thermodynamical linear response operational bounds.
%
%
We exemplified our approach with a quantum-mechanical full counting statistics 
description of a photoelectric device.
In multi-level systems it may be necessary to define multiple effective temperatures  for a non-canonical bath, 
corresponding to different transitions in the system. 
The identification of an effective temperature here and in other studies \cite{Yi, David1} was achieved in 
the limit of weak coupling between the qubit and the 
environment. Quantum systems that are {\it strongly} coupled to
equilibrium thermal reservoirs are expected to bring in new design rules 
for energy conversion devices \cite{DavidS, Kosloff,Cao,Gernot,Jarzynski,esposito17,Nazir}. 
Describing heat engines that are strongly coupled to non-canonical reservoirs remain a challenge for future work.



\begin{acknowledgments}
DS and BKA acknowledge support from an NSERC
Discovery Grant, the Canada Research Chair program,
and the CQIQC at the University of Toronto.
JHJ acknowledges supports from the National Science
Foundation of China (no. 11675116) and the Soochow university faculty start-up
funding. 
\end{acknowledgments}


\begin{thebibliography}{100}
\bibitem{comment}
Since $\langle N\rangle$ and $\langle Q_{ph}\rangle$ are extenstive properties that scale linearly with the time interval
$t$ in the steady state limit,
this definition of efficiency coincides with the definition that takes a ratio over the respective currents.
\end{thebibliography}
\end{document}